# Necessary and Sufficient Condition for Stability of Generalized Expectation Value

**Aziz El Kaabouchi[1] and Sumiyoshi Abe[1,2]**

[1] *Institut Supérieur des Matériaux et Mécaniques Avancés, 44 F. A. Bartholdi, 72000 Le Mans, France*

[2] *Department of Physical Engineering, Mie University, Mie 514-8507, Japan*

A class of generalized definitions of expectation value is often employed in nonequilibrium statistical mechanics for complex systems. Here, the necessary and sufficient condition is presented for such a class to be stable under small deformations of a given arbitrary probability distribution.

Given a probability distribution $\{p_i\}_{i=1,2,...,W}$, i.e., $0 \leq p_i \leq 1$ ($i=1,2,...,W$) and $\sum_{i=1}^{W} p_i = 1$, the ordinary expectation value of a quantity $Q$ of a system under consideration is defined by $\sum_{i=1}^{W} p_i Q_i$, where $W$ is the total number of accessible states and is enormously large in statistical mechanics, typically being $2^{10^{23}}$. In the field of generalized statistical mechanics for complex systems, on the other hand, discussions are often made about altering this definition. Among others, the so-called "escort average" is widely employed in the field of generalized statistical mechanics [1-3]. It is defined as follows:

$$\langle Q \rangle_\phi [p] = \sum_{i=1}^{W} P_i^{(\phi)} Q_i, \tag{1}$$

where $P_i^{(\phi)}$ stands for the escort probability distribution [4] given by

$$P_i^{(\phi)} = \frac{\phi(p_i)}{\sum_{j=1}^{W} \phi(p_j)}, \tag{2}$$

with a nonnegative function $\phi$. In the special case when $\phi(x) = x$, $\langle Q \rangle_\phi$ is reduced to the ordinary expectation value mentioned above.

Consider measurements of a certain quantity of a system to obtain information about the probability distribution. Repeated measurements should be performed on the system, which is identically prepared each time. Suppose that two probability distributions, $\{p_i\}_{i=1,2,...,W}$ and $\{p'_i\}_{i=1,2,...,W}$, are obtained through the measurements. They may slightly be different from each other, in general. If such measurements make sense, then the expectation values, $\langle Q \rangle [p]$ and $\langle Q \rangle [p']$, calculated from these two distributions should also be close to each other. This condition, which implies "experimental robustness", is represented as follows.

**Definition (Stability).** *An expectation value $\langle Q \rangle [p]$ is said to be stable, if the following predicate holds for any pair of probability distributions, $\{p_i\}_{i=1,2,...,W}$ and $\{p'_i\}_{i=1,2,...,W}$:*

$$(\forall \varepsilon > 0)(\exists \delta > 0)(\forall W)\left( \| p - p' \|_1 < \delta \Rightarrow \left| \langle Q \rangle [p] - \langle Q \rangle [p'] \right| < \varepsilon \right). \tag{3}$$

Here, $\| p - p' \|_1 = \sum_{i=1}^{W} |p_i - p'_i|$ is the $l^1$-norm describing the distance between these two probability distributions. One might consider norms of other kinds, but what is physically relevant to discrete systems is the present $l^1$-norm [5]. Equation (3) is analogous to Lesche's stability condition on entropic functionals [5], which has recently been revisited in the literature [6-11] (note that the discussion in Ref. [8] is corrected in Ref. [9]). This concept of stability is actually equivalent to that of uniform continuity.

In recent papers [12,13], it has been shown that the generalized expectation value in equation (1) with a specific class, $\phi(x) = x^q$ ($q > 0$), (the associated expectation value being termed the *q*-expectation value), is not stable unless $q = 1$. This result need the *q*-expectation-value formalism of nonextensive statistical mechanics [1,2] be reconsidered. In addition, the result is supported by Boltzmann-like kinetic theory in an independent manner [14].

Here, it seems appropriate to make some comments on the latest situation of the problems concerning stabilities of entropic functionals and generalized expectation values. The authors of Refs. [15,16] have presented discussions which aim to rescue the $q$-expectation values from the difficulties of their instability pointed out in Ref. [12]. Those authors insist that the $q$-expectation values can be stable in both the finite-$W$ and continuous cases. Such possibilities are however fully refuted by the work in Ref. [13] both physically and mathematically, and the controversy seems to have been terminated with that work. The case of the continuous variables was further been carefully examined in a recent paper [17], where the so-called Tsallis $q$-entropies [1,2] do not have the continuous limit in consistency with the physical principles such as the thermodynamic laws (see also [18,19]). These controversies have led the researchers to give up the traditional form of nonextensive statistical mechanics based on the $q$-entropies and $q$-expectation values and to examine other entropic functionals combined with the ordinary definition of expectation values [20] (see also Refs. [21,22]). Thus, it seems that nonextensive statistical mechanics has to be fully reexamined, theoretically.

In this paper, we present the necessary and sufficient condition for $\langle Q \rangle_\phi [p]$ in equation (1) to be stable. This result is of obvious importance for generalized statistical mechanics for complex systems.

Our main result is as follows.

**Theorem.** *Let $\phi$ be nonnegative and continuous on $[0,1]$, differentiable on $(0,1)$, and satisfy the condition that $\phi(x) = 0 \Leftrightarrow x = 0$. And, let $Q = \{Q_i\}_{i=1,2,...,W}$ be a* random variable. Then, $\langle Q \rangle_\phi [p]$ in equation (1) is stable, if and only if $\lim_{x \to +0} \phi(x)/x \in (0, \infty)$.

*Proof.* First, assume that $\lim_{x \to +0} \phi(x)/x = a > 0$. Then, there exists $\delta_1 > 0$ such that

$$a - \frac{a}{2} < \frac{\phi(x)}{x} < a + \frac{a}{2} \qquad (\forall x \in (0, \delta_1]). \tag{4}$$

$\phi(x)/x$ does not vanish because of the condition $\phi(x) = 0 \Leftrightarrow x = 0$. Therefore, there exists $b > 0$ such that

$$\frac{\phi(x)}{x} \geq b \qquad (\forall x \in (\delta_1, 1]). \tag{5}$$

Putting $c = \min\{a/2, b\}$, we have

$$c x \leq \phi(x) \qquad (\forall x \in [0, 1]). \tag{6}$$

Consequently, for an arbitrarily large $W$ and an arbitrary probability distribution $\{p_i\}_{i=1, 2, \ldots, W}$, we obtain

$$\frac{1}{\sum_{i=1}^{W} \phi(p_i)} \leq c. \tag{7}$$

From the mean value theorem, it follows that

$$\left| \phi(p_i) - \phi(p'_i) \right| \leq \left| p_i - p'_i \right| \cdot \sup_{x \in (0,1)} \left| \phi'(x) \right|, \tag{8}$$

where $\phi'(x)$ is the derivative of $\phi(x)$ with respect to $x$. For $\varepsilon > 0$, we put

$$\delta = \inf\left(\delta_1, \frac{c\varepsilon}{2|Q_{\max}| \cdot \left(\sup_{x \in (0,1)} |\phi'(x)|\right)}\right), \tag{9}$$

where $Q_{\max} = \max\{Q_i\}_{i=1,2,\ldots,W}$. Now, for $\|p - p'\|_1 < \delta$, we have

$$\left|\langle Q \rangle_\phi[p] - \langle Q \rangle_\phi[p']\right|$$

$$= \frac{1}{\sum_{i=1}^{W} \phi(p_i) \sum_{j=1}^{W} \phi(p'_j)} \left|\sum_{i=1}^{W} Q_i \left\{\phi(p_i) \sum_{j=1}^{W} \phi(p'_j) - \phi(p'_i) \sum_{j=1}^{W} \phi(p_j)\right\}\right|$$

$$\leq \frac{1}{\sum_{i=1}^{W} \phi(p_i) \sum_{j=1}^{W} \phi(p'_j)}$$

$$\times \left[\sum_{i=1}^{W} |Q_i| \left\{|\phi(p_i) - \phi(p'_i)| \sum_{j=1}^{W} \phi(p'_j) + \phi(p'_i) \left|\sum_{j=1}^{W} \phi(p_j) - \sum_{j=1}^{W} \phi(p'_j)\right|\right\}\right]$$

$$\leq \frac{1}{\sum_{j=1}^{W} \phi(p_j)} \sum_{i=1}^{W} |Q_i| |\phi(p_i) - \phi(p'_i)|$$

$$+ \frac{\sum_{j=1}^{W} |\phi(p_j) - \phi(p'_j)|}{\sum_{i=1}^{W} \phi(p_i) \sum_{j=1}^{W} \phi(p'_j)} \sum_{i=1}^{W} |Q_i| \phi(p'_i)$$

$$\leq \frac{2|Q_{max}|}{\sum_{j=1}^{W}\phi(p_j)}\sum_{i=1}^{W}|\phi(p_i)-\phi(p'_i)|,$$

$$\leq \frac{2|Q_{max}|}{\sum_{j=1}^{W}\phi(p_j)}\|p-p'\|_1 \cdot \sup_{x\in(0,1)}|\phi'(x)|$$

$$\leq \frac{2|Q_{max}|}{c}\|p-p'\|_1 \cdot \sup_{x\in(0,1)}|\phi'(x)|$$

$$< \varepsilon. \tag{10}$$

Therefore, $\langle Q \rangle_\phi [p]$ is stable.

On the other hand, suppose that $\lim_{x\to+0}\phi(x)/x \notin (0,\infty)$. That is, (i) $\lim_{x\to+0}\phi(x)/x = 0$ or (ii) $\lim_{x\to+0}\phi(x)/x = \infty$. Below, we shall examine these cases separately.

(i) Consider the following deformation:

$$p_i = \frac{1}{W-1}(1-\delta_{i1}), \tag{11}$$

$$p'_i = \left(1-\frac{\delta}{2}\right)p_i + \frac{\delta}{2}\delta_{i1}, \tag{12}$$

which are normalized and satisfy $\|p-p'\|_1 = \delta$. We have

$$\sum_{i=1}^{W}\phi(p_i) = (W-1)\phi\left(\frac{1}{W-1}\right), \tag{13}$$

$$\sum_{i=1}^{W}\phi(p'_i)=\phi\left(\frac{\delta}{2}\right)+(W-1)\phi\left(\frac{1}{W-1}\left(1-\frac{\delta}{2}\right)\right). \tag{14}$$

Difference of the expectation values is calculated as follows:

$$\langle Q\rangle_\phi[p]-\langle Q\rangle_\phi[p']$$

$$=-\frac{Q_1\phi\left(\frac{\delta}{2}\right)}{\phi\left(\frac{\delta}{2}\right)+(W-1)\phi\left(\frac{1}{W-1}\left(1-\frac{\delta}{2}\right)\right)}$$

$$+\left(\sum_{i=2}^{W}Q_i\right)\left\{\frac{1}{W-1}-\frac{\phi\left(\frac{1}{W-1}\left(1-\frac{\delta}{2}\right)\right)}{\phi\left(\frac{\delta}{2}\right)+(W-1)\phi\left(\frac{1}{W-1}\left(1-\frac{\delta}{2}\right)\right)}\right\}$$

$$=\frac{W}{W-1}(\bar{Q}-Q_1)\frac{\phi\left(\frac{\delta}{2}\right)/\left(1-\frac{\delta}{2}\right)}{\phi\left(\frac{\delta}{2}\right)/\left(1-\frac{\delta}{2}\right)+\phi\left(\frac{1}{W-1}\left(1-\frac{\delta}{2}\right)\right)/\left[\frac{1}{W-1}\left(1-\frac{\delta}{2}\right)\right]}$$

$$\xrightarrow{W\to\infty}\bar{Q}-Q_1, \tag{15}$$

since $\lim_{x\to+0}\phi(x)/x=0$, where $\bar{Q}$ is the arithmetic mean, $\bar{Q}=\sum_{i=1}^{W}Q_i/W$. Therefore, $\langle Q\rangle_\phi[p]$ is not stable.

(ii) Consider the following deformation:

$$p_i=\delta_{i1}, \tag{16}$$

$$p'_i = \left(1 - \frac{\delta}{2}\frac{W}{W-1}\right)p_i + \frac{\delta}{2}\frac{1}{W-1}, \tag{17}$$

which are also normalized and satisfy $\|p - p'\|_1 = \delta$. We have

$$\sum_{i=1}^{W} \phi(p_i) = \phi(1), \tag{18}$$

$$\sum_{i=1}^{W} \phi(p'_i) = \phi\left(1 - \frac{\delta}{2}\right) + (W-1)\phi\left(\frac{\delta}{2}\frac{1}{W-1}\right). \tag{19}$$

Difference of the expectation values is calculated as follows:

$$\langle Q \rangle_\phi [p] - \langle Q \rangle_\phi [p']$$

$$= Q_1 \left\{ 1 - \frac{\phi\left(1 - \frac{\delta}{2}\right)}{\phi\left(1 - \frac{\delta}{2}\right) + (W-1)\phi\left(\frac{\delta}{2}\frac{1}{W-1}\right)} \right\}$$

$$- \left(\sum_{i=2}^{W} Q_i\right) \frac{\phi\left(\frac{\delta}{2}\frac{1}{W-1}\right)}{\phi\left(1 - \frac{\delta}{2}\right) + (W-1)\phi\left(\frac{\delta}{2}\frac{1}{W-1}\right)}$$

$$= \frac{W}{W-1}(Q_1 - \bar{Q}) \frac{\phi\left(\frac{\delta}{2}\frac{1}{W-1}\right) / \left[\frac{\delta}{2}\frac{1}{W-1}\right]}{\phi\left(1 - \frac{\delta}{2}\right) / \left(\frac{\delta}{2}\right) + \phi\left(\frac{\delta}{2}\frac{1}{W-1}\right) / \left[\frac{\delta}{2}\frac{1}{W-1}\right]}$$

$$\xrightarrow{W \to \infty} Q_1 - \bar{Q}, \tag{20}$$

since $\lim_{x \to +0} \phi(x)/x = \infty$. Therefore, $\langle Q \rangle_\phi [p]$ is not stable. □

In the above proof, we have employed the specific deformations of the probability distributions as the counterexamples, which are considered in Ref. [5]. It is pointed out in Ref. [13] that these deformed distributions may experimentally be generated.

Finally, we mention a couple of simple stable examples.

Example 1:

$$\phi(x) = e^x - 1. \tag{22}$$

Example 2:

$$\phi(x) = \ln(1 + x^\alpha), \tag{23}$$

which yields a stable generalized expectation value, if and only if $\alpha = 1$.

On the other hand, as mentioned earlier, the $q$-expectation value is not stable, since $\phi(x) = x^q$ ($q > 0, q \neq 1$) does not satisfy the condition $\lim_{x \to +0} \phi(x)/x \in (0, \infty)$.

In conclusion, we have considered a class of generalized definitions of expectation value that are often employed in nonequilibrium statistical mechanics for complex systems, and have presented the necessary and sufficient condition for such a class to be stable under small deformations of a given arbitrary probability distribution.

**Acknowledgment**

The work of S.A. was supported in part by a Grant-in-Aid for Scientific Research from the Japan Society for the Promotion of Science.